\documentstyle[12pt,amssymb]{amsart}
\newmathalphabet*{\frakb}{euf}{b}{n}
\newmathalphabet*{\eusb}{eus}{b}{n}
\newmathalphabet*{\eur}{eur}{m}{n}
\size{10}{18pt}\selectfont
\begin{document}

\title{Oracle Complexity and Nontransitivity in Pattern Recognition}
\author{Vadim Bulitko \\ bulitko@@ualberta.ca }

%\vspace*{-2.0cm}

\date{Written in Spring 1995}

\maketitle

\hyphenation{orac-le}

\begin{abstract}
Different mathematical models of recognition processes are known. In pre\-sent
paper we consider a recognizing algorithm as an oracle computation on Turing
machine \cite{Rog}. Such point of view seems to be useful in pattern
recognition as well as in recursion theory. Use of recursion theory in
pattern recognition shows connection between a recognition algorithm
comparison problem and complexity problems of oracle computation \cite{Bul}.
That is because in many cases we can take into account only the number of
sign computations or in other words volume of oracle information needed.
Therefore, the problem of recognition algorithm preference can be formulated
as a complexity optimization problem of oracle computation. Furthermore,
introducing a certain "natural" preference relation on a set of
recognition algorithms we discover it to be nontransitive. This relates to the
well known nontransitivity paradox in probability theory \cite{Sz}.


\end{abstract}

Most of our notation follows \cite{Rog}, also r.e. means recursively
enumerable.

\smallskip

Let $\left( P_i\right) _{i\in \Bbb N}$ be a system of one-place predicates (basic
signs) defined on a pattern set $U$. We assume $(\forall a,b\in U)\left[
a\neq b\Rightarrow \exists i\left[ P_i(a)\neq P_i(b)\right] \right] $. Thus
it is possible to represent every pattern $a\in U$ as a subset $\widetilde{a}%
\subset \Bbb N:$ $\widetilde{a}=\left\{ i\mid P_i(a)\right\} $. Also $\widehat{D}%
=a\rightleftharpoons \widetilde{a}=D$. So an image $B$, which is a set of
patterns from $U$, can be represented as a subset of $2^{\Bbb N}$. Let $S$ be a set
of all images. So the recognition problem for given, in general partial,
function $\tau :\Bbb N\rightarrow S$ and given pattern set $U$ can be formulated
as to find such $z$ that for every pattern $x\in U$:

\begin{center}
$\varphi _z^{\widetilde{x}}\left( i\right) =\left\{ 
\begin{array}{l}
1, 
\text{ if }x\in \tau \left( i\right) \text{and }\tau \left( i\right) \text{%
is defined} \\ 0, 
\text{ if }x\notin \tau \left( i\right) \text{and }\tau \left( i\right) 
\text{is defined} \\ \text{undefined},\text{if }\tau \left( i\right) \text{%
is undefined} 
\end{array}
\right. $
\end{center}

If such $z$ exists then we say the recognition problem is solvable for
triple $(U,\tau ,$ $\left( P_i\right) _{i\in \Bbb N})$.

{\sl Definition.} {\it We use $U^f$ for $U$ if \begin{enumerate}
\item $\forall x\in U\;[\left| 
\widetilde{x}\right| <\infty ]$; 
\item $\{p\mid D_p=\widetilde{x},x\in U\}$
is recursive.
\end{enumerate} }

{\sl Definition.} {\it $\tau $ is called computable $\rightleftharpoons $ $%
\left( \exists \text{ partial recursive }\gamma \right) $ $\forall i$} {\it $%
[$}$( \tau (i)\downarrow \Leftrightarrow%\linebreak
\gamma (i)\downarrow ) $ 
$\&$ $( \gamma (i)\downarrow \Rightarrow W_{\gamma (i)}=\{p\mid D_p= 
\widetilde{x},x\in \tau (i)\}) ]${\it .}

The following proposition can be easily proved.

{\sl Proposition.} \begin{enumerate}
\item {\it There is a triple $\left( U,\tau ,\left(
P_i\right) _{i\in \Bbb N}\right) $ such that $\tau $} {\it is total,} {\it $\tau
(i)$ is countable for every $i$},{\it \ and the recognition problem for $%
\left( U,\tau ,\left( P_i\right) _{i\in \Bbb N}\right) $ is unsolvable. }

\item {\it For given non-empty set $U^f$ there is a computable total $\tau $
such that the recognition problem for $\left( U^f,\tau ,\left( P_i\right)
_{i\in \Bbb N}\right) $ is unsolvable}.
\end{enumerate}

We should consider unsolvability of a recognition problem as an indication
of non-adequacy of the basic sign system to image set{\it \ $S$} and
function {\it $\tau $}. This means any algorithm $z$ makes mistakes. However
some of the algorithms might make more mistakes than others. The following
is a formalization of that.

Let us consider {\it $\left( U^f,\tau ,\left( P_i\right) _{i\in \Bbb N}\right) $}%
, where $\tau $ is computable. 
Define ${\cal B}=\{\left\langle
k,i\right\rangle \mid k\in W_{\gamma (i)}$ (i.e $\widehat{D}_k\in \tau (i)$)$%
\}$. ${\cal B}$ is r.e. If ${\cal B}$ is not recursive then {\it $\left(
U^f,\tau ,\left( P_i\right) _{i\in \Bbb N}\right) $} is unsolvable, however every
algorithm $z$ has a finite {\it correctness domain} $\left[ 0,n\right] $
equal to such a maximal initial segment of $\Bbb N$ that $\left( \forall j\in
\left[ 0,n\right] \right) \{[ j=\left\langle k,i\right\rangle \in {\cal %
B}\Rightarrow \varphi _z^{D_k}(i)=1] $ $\&$ $\left[ j=\left\langle
k,i\right\rangle \notin {\cal B}\Rightarrow \varphi _z^{D_k}(i)=0\right] \}$.

So it is natural to choose new signs on patterns from $U^f$ such that there
is new recognition algorithm $\varphi _{z^{\prime }}$ with correctness
domain strictly including correctness domain of $\varphi _z$. Definitely we
have to keep doing that all the time and use new information about the
images. It is possible to estimate the volume of the information using its
Kolmogorov complexity \cite{Lov}. Namely given computable $\tau $ the volume
of information sufficient to define new signs $P_i^{*}$ and construct
recognition algorithm for $\left( U^f,\tau ,\left( P_i^{*}\right) _{i\in
\Bbb N}\right) $ ,that has correctness domain with length $n$, grows not faster
than $\ln n$.

Usually an image includes a given pattern along with its various
modifications. They can be got by effective transformations of the pattern.
In that more general case let ${\cal B}=( \underset{q\in \Theta }
{\bigcup }R_q) \bigcap W_e$ where $\Theta \subset \Bbb N$, $R_q$ is a
recursive set such that $C_{R_q}=\varphi _{g(q)}$; and $\min R_q\geq h(q)$
for every $q\in \Theta $; here $g,h$ are total recursive functions and $h$
is not a decreasing function. $KR_\Theta $ is resolving complexity of $%
\Theta $ \cite{Lov}.

{\sl Proposition.} {\it Taking into account above made assumptions we do not
need more than $\ln n+KR_\Theta (h^{-1}(n))$ of information about $W_e\oplus
\Theta $ to construct new signs $P_n^{*}$ such that the recognition problem $%
( U^f, ( \tau (i_0)) ,( P_n^{*}) _{n\in \Bbb N}) $
is solvable.}

In a training process a hypothesis about teacher's signs can be transformed
into a hypothesis about enumeration operator. So a recognition algorithm can
be found using the Kleene recursion theorem. The last proposition allows to
estimate frequency and volume of operator hypothesis changes.

So we can introduce time of the training. It would be interesting to
introduce time in a recognition process. We propose the following method to
do that.

As well known, there is a passage from r.e. set $W_z$ defining partial
recursive in $A$ function $\varphi _z^A$ to tree diagram $T_z$ of the Turing
machine computation with oracle $A$. Tree $T_z$ can be equivalently
represented as r.e. set $W_{f(z)}$ containing the tree's terminated
branches. A passage from $W_z$ to $W_{f(z)}$ is called regularization. It
can be done in such a way that $f$ will be recursive and some additional
conditions for $W_{f(z)}$ will be satisfied. Notation $\left\langle
x,y,u,v\right\rangle \in ^1W_{f(z)}$ means there is a branch $\left(
x,n_1\sigma _1n_2\sigma _2\ldots n_k\sigma _k,y\right) $ belonging to $%
W_{f(z)}$ such that $D_u=\left\{ n_t\mid \sigma _t=1\right\} \&$ $%
D_v=\left\{ n_t\mid \sigma _t=0\right\} $. $W_{z,t}$ denotes result of
enumeration of $W_z$ by algorithm $z$ after $t$ steps done.

{\sl Proposition.} {\it There is a recursive function $f$ such that} 
$\varphi _i^X(x)=y\Rightarrow \varphi _{f^1(z)}^X(x)%\linebreak
=y;$\\
$\forall x\forall \;\; y
\left[ \underset{\left\langle x,y,u,v\right\rangle \in Wz,t}{\bigcup }
\left( D_u\cup D_v\right) \supset \underset{\left\langle
x,y,u,v\right\rangle \in Wf^1(z),t}{\bigcup }\left( D_u\cup D_v\right)
\right]  \text{\it where } W_{f^1(z)} = %\linebreak 
\{\left\langle x,y,u,v
\right\rangle \mid \left\langle x,y,u,v\right\rangle \in ^1W_{f(z)}\}.$

So using this approach we can represent recognition algorithms as trees.
Vertexes of these trees contain the signs to be computed. The end of a
terminated branch of a tree contains $\varphi _z^{\widetilde{A}}\left(
i\right) $ (i.e. it indicates if $A\in \tau \left( i\right) $ or not).

In practice image system $S$ and patterns are finite and every $\eur{ a}_i\in
S$ is a recursive subset of $U^f$. In that case the recognition problem for $%
( U^f,\left( \eur{ a}_1,\ldots ,\eur{ a}_k\right) ,%\linebreak \left( P_n\right)
_{n\in \Bbb N}) $ is definitely solvable however there is a complexity
problem. Also we should mention the best algorithm problem comes up if we
introduce a certain preference relation on the set of recognition
algorithms. In order to consider that we assume sign computation is an easy
process (i.e. any sign computation takes one time unit).

Assume $\eur{ a}_i\cap \eur{ a}_j=\emptyset $ if $i\neq j$; $i,j\in \left\{
1,\ldots ,k\right\} $ and $U^f=\underset{i}{\cup }\eur{ a}_i$. Then
evidently there is a recursive function $q$ such that $\left( \forall x\in
U^f\right) \left[ \varphi _{q(z)}^x(0)=i\Leftrightarrow \varphi
_z^x(i)=1\right] $. Under these assumptions one tree represents one
recognition algorithm and any branch of such tree ends.

We call branch length (quantity of arcs) recognition time for pattern $x$ by
algorithm $A$ and write it $T\left( A,x\right) $. Recognition time for a set 
$M\subset U^f$ by algorithm $A$ is $T\left( A,M\right) =\max\limits_{x\in M}
T\left( A,x\right) $.

Let $R$ be a recognition algorithm set. Algorithm $A$ defeats algorithm $B$
recognition pattern $x$ iff $T\left( A,x\right) <T\left( B,x\right) $.
Define $V\left( A,B\right) = | \{ x\in U\mid T ( A,x)
< T(B,x) \} | $.

Consider a situation when algorithms $A$ and $B$ are in process of 
recognizing random
pattern sequences. Let $\pi (t),1\leq t\leq n$ be a random pattern sequence.
Let $v\left( x,t\right) $ be the probability of event $\pi \left( t\right) =x$.
So the mathematical expectation $m\left( A,B,n\right) $ of the number of
members in $\pi $ on which $A$ defeats $B$ can be expressed as follows:

\begin{center}
$\underset{x\in U^f}{\sum }\stackrel{n}{\underset{t=1}{\sum}}
v\left( x,t\right) sg\left( T\left( B,x\right) \bumpeq T\left(
A,x\right) \right) $
\end{center}

where $sg\left( a\right) =\left\{ 
\begin{array}{l}
1,a>0 \\ 
0,else 
\end{array}
\right. $, $a \bumpeq b=\left\{ 
\begin{array}{l}
a-b,a\geq b \\ 
0,else 
\end{array}
\right. $.

Assuming $v\left( x,t\right) =const$ we get $\left( \forall n\geq 1\right)
\left[ \frac{m(A,B,n)}n=\frac{V(A,B)}{\left| U^f\right| }\right] $ and
therefore recognizing the random pattern sequence algorithm $A$ wins more
times in average than algorithm $B$ iff $V\left( A,B\right) >V\left(
B,A\right) $. So we can introduce a preference relation on $R$ : we say
algorithm $A$ is better than algorithm $B$ (write $A\ll B$) iff $V\left(
A,B\right) >V\left( B,A\right) $. We call algorithm $A$ equivalent to $B$
iff $V\left( A,B\right) =V\left( B,A\right) $. It is clear that any two
algorithms are comparable and the preference relation seems to be natural.
Consider an example demonstrating nontransitivity of the introduced
preference relation.

Let us define $\frak{B}=\left\{ 0,1\right\} $, so $\frak{B}^9$ $=$ $\left\{
i_1i_2\ldots i_3\mid i_s\in \frak{B}\eur{ ,\ }s\in \overline{1,9}\right\} $ and
$\{i_1\ldots \linebreak i_{m_1-1}\frak{B}i_{m_1+1} \ldots i_{m_k-1}\frak{B}%
i_{m_k+1}\ldots i_9\}$ means $\{ i_1\ldots
i_{m_1-1}i_{m_1m_1+1}\ldots i_{m_k-1}i_{m_k} %\linebreak
i_{m_k+1}\ldots i_9\mid
i_{m_s}\in \frak{B}; s=\overline{1,k};0\leq k\leq 9\} \subset \frak{B}^9$.
Here and below $\overline{a,b}$ means $\left\{ a,\ldots ,b\right\} $.

Now define $U^f\subset \frak{B}^9$ as $U^f=\alpha _0\cup \alpha _1\cup \alpha
_2\cup \alpha _3$ where $\alpha _0=\{1\frak{BB}01\frak{B}001\}$, $\alpha
_1=\{01\frak{B}0011\frak{BB}\}$, $\alpha _2=\{0011\frak{BB}01\frak{B}\}$, $%
\alpha _3=\{000000000\}$. Obviously $\left| \alpha _0\right| =\left| \alpha
_1\right| =\left| \alpha _2\right| =8,\left| \alpha _3\right| =1,\left|
U^f\right| =25$. Set $S=\{\alpha _0,\alpha _1,\alpha _2,\alpha _3\}$.

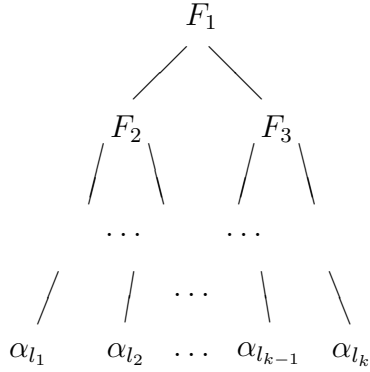
\begin{figure}
\begin{center}
\unitlength=1.00mm
\special{em:linewidth 0.4pt}
\linethickness{0.4pt}
\begin{picture}(47.00,47.00)
\put(28.00,43.00){\line(1,-1){7.00}}
\put(25.00,43.00){\line(-1,-1){7.00}}
\put(40.00,30.00){\line(1,-4){2.00}}
\put(34.00,30.00){\line(-1,-4){2.00}}
\put(20.00,30.00){\line(1,-4){2.00}}
\put(14.00,30.00){\line(-1,-4){2.00}}
\put(8.00,13.00){\line(-2,-5){2.78}}
\put(18.00,13.00){\line(-1,-6){1.11}}
\put(35.00,13.00){\line(1,-6){1.11}}
\put(44.00,13.00){\line(2,-5){2.78}}
\put(27.00,47.00){\makebox(0,0)[cc]{$F_1$}}
\put(17.00,32.00){\makebox(0,0)[cc]{$F_2$}}
\put(37.00,32.00){\makebox(0,0)[cc]{$F_3$}}
\put(17.00,18.00){\makebox(0,0)[cc]{$\dots$}}
\put(33.00,18.00){\makebox(0,0)[cc]{$\dots$}}
\put(26.00,10.00){\makebox(0,0)[cc]{$\dots$}}
\put(4.00,2.00){\makebox(0,0)[cc]{$\alpha_{l_1}$}}
\put(17.00,2.00){\makebox(0,0)[cc]{$\alpha_{l_2}$}}
\put(26.00,2.00){\makebox(0,0)[cc]{$\dots$}}
\put(36.00,2.00){\makebox(0,0)[cc]{$\alpha_{l_{k-1}}$}}
\put(47.00,2.00){\makebox(0,0)[cc]{$\alpha_{l_k}$}}
\end{picture}
\caption{\it  Structure of recognition algorithms}
\label{fig1}
\end{center}
\end{figure}

Consider nine predicates on $U^f{\cal \ }$defined as $P_k\left( x\right)
=i_k,k=\overline{1,9}$. Also we consider a set ${\cal R}$ of recognition
algorithms which are binary trees in the form shown on [Fig. \ref{fig1}].

There $F_i\in \left\{ P_k\mid k=\overline{1,9}\right\} ,\alpha _{l_q}\in
\left\{ \alpha _s\mid s=\overline{0,3}\right\} $. We call $F_1$ the first
predicate in the algorithm or predicate in the root.

Every pattern $x$ corresponds to a branch in tree of algorithm $A$ from the
root to a leaf. The leaf is output of algorithm $A$ recognizing pattern $x$.
We assume jump to the left subnode of node $F_i$ iff $F_i\left( x\right) $
is true.

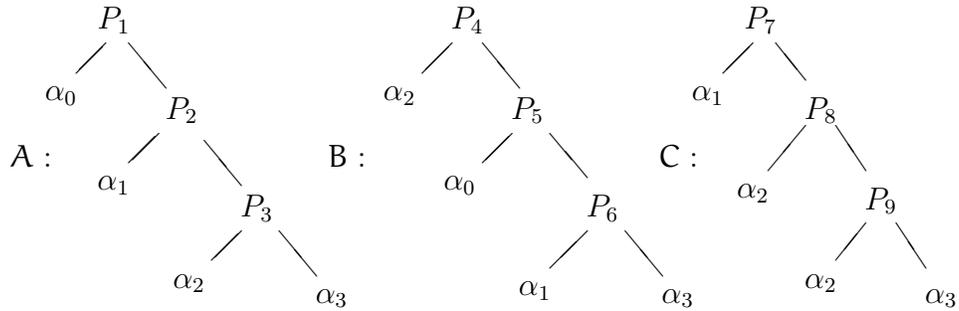
\begin{figure}
\begin{center}
\unitlength=1.00mm
\special{em:linewidth 0.4pt}
\linethickness{0.4pt}
\begin{picture}(121.00,48.00)
\put(13.00,40.00){\line(5,-6){5.00}}
\put(23.00,27.00){\line(5,-6){5.00}}
\put(33.00,15.00){\line(5,-6){5.00}}
\put(28.00,15.00){\line(-1,-1){4.00}}
\put(60.00,40.00){\line(5,-6){5.00}}
\put(64.00,28.00){\line(-1,-1){4.00}}
\put(69.00,28.00){\line(5,-6){5.00}}
\put(79.00,15.00){\line(5,-6){5.00}}
\put(99.00,40.00){\line(4,-5){4.00}}
\put(107.00,29.00){\line(2,-3){4.00}}
\put(115.00,16.00){\line(2,-3){4.00}}
\put(111.00,16.00){\line(-4,-5){4.00}}
\put(103.00,29.00){\line(-5,-6){5.00}}
\put(10.00,40.00){\line(-1,-1){4.00}}
\put(56.00,40.00){\line(-1,-1){4.00}}
\put(96.00,40.00){\line(-1,-1){4.00}}
\put(58.00,43.00){\makebox(0,0)[cc]{$P_4$}}
\put(97.00,43.00){\makebox(0,0)[cc]{$P_7$}}
\put(11.00,43.00){\makebox(0,0)[cc]{$P_1$}}
\put(20.00,31.00){\makebox(0,0)[cc]{$P_2$}}
\put(66.00,31.00){\makebox(0,0)[cc]{$P_5$}}
\put(105.00,31.00){\makebox(0,0)[cc]{$P_8$}}
\put(30.00,18.00){\makebox(0,0)[cc]{$P_3$}}
\put(74.00,15.00){\line(-1,-1){5.00}}
\put(76.00,18.00){\makebox(0,0)[cc]{$P_6$}}
\put(113.00,19.00){\makebox(0,0)[cc]{$P_9$}}
\put(4.00,33.00){\makebox(0,0)[cc]{$\alpha_0$}}
\put(11.00,21.00){\makebox(0,0)[cc]{$\alpha_1$}}
\put(21.00,8.00){\makebox(0,0)[cc]{$\alpha_2$}}
\put(40.00,6.00){\makebox(0,0)[cc]{$\alpha_3$}}
\put(49.00,33.00){\makebox(0,0)[cc]{$\alpha_2$}}
\put(57.00,21.00){\makebox(0,0)[cc]{$\alpha_0$}}
\put(67.00,7.00){\makebox(0,0)[cc]{$\alpha_1$}}
\put(86.00,6.00){\makebox(0,0)[cc]{$\alpha_3$}}
\put(90.00,33.00){\makebox(0,0)[cc]{$\alpha_1$}}
\put(96.00,20.00){\makebox(0,0)[cc]{$\alpha_2$}}
\put(105.00,8.00){\makebox(0,0)[cc]{$\alpha_2$}}
\put(121.00,6.00){\makebox(0,0)[cc]{$\alpha_3$}}
\put(0.00,25.00){\makebox(0,0)[cc]{$\eur{A}:$}}
\put(42.00,25.00){\makebox(0,0)[cc]{$\eur{B}:$}}
\put(86.00,25.00){\makebox(0,0)[cc]{$\eur{C}:$}}
\put(13.00,24.00){\line(1,1){4.00}}
\end{picture}
\caption{\it  Recognition algorithms $\eur{A},\eur{B},\eur{C}$ }
\label{fig2}
\end{center}
\end{figure}

Finally consider three algorithms shown on [Fig. \ref{fig2}].

[Tab. \ref{tab1}] shows their recognition time.

\begin{table}
\begin{center}
\begin{tabular}{|c||c|c|c|}
\hline
$M\backslash Y$ & $\eur{A}$ & $\eur{B}$ & $\eur{C}$ \\ 
\hline
$\alpha _0$ & 1 & 2 & 3 \\ 
\hline
$\alpha _1$ & 2 & 3 & 1 \\ 
\hline
$\alpha _2$ & 3 & 1 & 2 \\ 
\hline
$\alpha _3$ & 3 & 3 & 3 \\
\hline
\end{tabular}
\caption{\it Recognition time of algorithms $\eur{A},\eur{B},\eur{C}$ }
\label{tab1}
\end{center}
\end{table}

{\sl Theorem 1}. \begin{enumerate}
\item $\eur{A}\ll \eur{B},\eur{B}\ll \eur{C},\eur{C}\ll 
\eur{A}$. {\it Therefore the preference relation is nontransitive.}
\item {\it There is no such} ${\cal X}\in {\cal R}\setminus \left\{ \eur{A},%
\eur{B},\eur{C}\right\} $ {\it that} $\left( \eur{A}\ll {\cal X}\right) \vee
\left( \eur{B}\ll {\cal X}\right) \vee \left( \eur{C}\ll {\cal X}\right) $.
\end{enumerate}

{\sl Proof.}

The first statement follows from [Tab. \ref{tab1}].

Consider the second statement. Let ${\cal X}$ be an arbitrary recognition
algorithm. We have to investigate two cases: \begin{enumerate}
\item $\forall x\in U^f\left[T\left( {\cal X},x\right) \geq 2\right] $; 
\item $\exists x\in U^f\left[T\left( {\cal X},x\right) =1\right] $.
\end{enumerate}

For the {\it first case} we will go through all the predicates which can be
situated in the root of the tree of ${\cal X}$. Those predicates are $P_2$, $%
P_3$, $P_5$, $P_6$, $P_8$, $P_9$ because $P_1$, $P_4$, $P_7$ separate an
entire image. We will pay attention only to $P_2$ because the other cases
are similar.

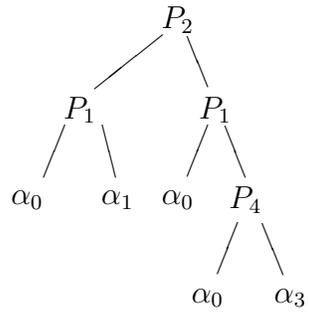
\begin{figure}
\begin{center}
\unitlength=1.00mm
\special{em:linewidth 0.4pt}
\linethickness{0.4pt}
\begin{picture}(40.00,39.00)
\put(39.00,5.00){\line(-2,5){2.74}}
\put(33.00,12.00){\line(-2,-5){2.73}}
\put(34.00,18.00){\line(-2,5){2.73}}
\put(29.00,25.00){\line(-2,-5){2.72}}
\put(29.00,30.00){\line(-2,5){2.72}}
\put(23.00,37.00){\line(-5,-4){9.03}}
\put(10.00,25.00){\line(-2,-5){2.79}}
\put(15.00,25.00){\line(1,-4){1.74}}
\put(12.00,27.00){\makebox(0,0)[cc]{$P_1$}}
\put(30.00,27.00){\makebox(0,0)[cc]{$P_1$}}
\put(5.00,15.00){\makebox(0,0)[cc]{$\alpha_0$}}
\put(17.00,15.00){\makebox(0,0)[cc]{$\alpha_1$}}
\put(25.00,15.00){\makebox(0,0)[cc]{$\alpha_0$}}
\put(34.00,15.00){\makebox(0,0)[cc]{$P_4$}}
\put(29.00,2.00){\makebox(0,0)[cc]{$\alpha_0$}}
\put(40.00,2.00){\makebox(0,0)[cc]{$\alpha_3$}}
\put(25.00,39.00){\makebox(0,0)[cc]{$P_2$}}
\end{picture}
\caption{\it The first sub case}
\label{fig3}
\end{center}
\end{figure}

So suppose $P_2$ is the first predicate computed in ${\cal X}$. It will
''divide'' all the patterns into two sets: $\left\{ \frac 12\alpha _0,\alpha
_1\right\} $ and $\left\{ \frac 12\alpha _0,\alpha _2,\alpha _3\right\} $.
Here and below $\frac pq\alpha _0$ is a subset of $\alpha _0$ such that $%
\left| \frac pq\alpha _0\right| =\frac pq\left| \alpha _0\right| $. It is
clear we can easily split the first of those two sets. However we cannot
split the second set by one predicate because the second set contains
patterns belonging to the three images. So we get two sub cases:

(1) we separate patterns belonging to $\alpha _0$ from the second set and
split $\alpha _2,\alpha _3$ by another predicate. The scheme on 
[Fig. \ref{fig3}] shows that sub case.

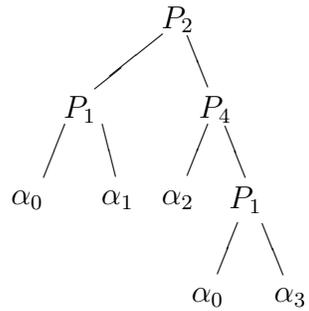
\begin{figure}
\begin{center}
\unitlength=1.00mm
\special{em:linewidth 0.4pt}
\linethickness{0.4pt}
\begin{picture}(40.00,39.00)
\put(39.00,5.00){\line(-2,5){2.74}}
\put(33.00,12.00){\line(-2,-5){2.73}}
\put(34.00,18.00){\line(-2,5){2.73}}
\put(29.00,25.00){\line(-2,-5){2.72}}
\put(29.00,30.00){\line(-2,5){2.72}}
\put(23.00,37.00){\line(-5,-4){9.03}}
\put(10.00,25.00){\line(-2,-5){2.79}}
\put(15.00,25.00){\line(1,-4){1.74}}
\put(12.00,27.00){\makebox(0,0)[cc]{$P_1$}}
\put(30.00,27.00){\makebox(0,0)[cc]{$P_4$}}
\put(5.00,15.00){\makebox(0,0)[cc]{$\alpha_0$}}
\put(17.00,15.00){\makebox(0,0)[cc]{$\alpha_1$}}
\put(25.00,15.00){\makebox(0,0)[cc]{$\alpha_2$}}
\put(34.00,15.00){\makebox(0,0)[cc]{$P_1$}}
\put(29.00,2.00){\makebox(0,0)[cc]{$\alpha_0$}}
\put(40.00,2.00){\makebox(0,0)[cc]{$\alpha_3$}}
\put(25.00,39.00){\makebox(0,0)[cc]{$P_2$}}
\end{picture}
\caption{\it  The second sub case}
\label{fig4}
\end{center}
\end{figure}

(2) we separate $\alpha _2$ from the second set first and then split $\frac
12\alpha _0$ and $\alpha _3$ [see Fig. \ref{fig4}].

However comparison of these algorithms with $\eur{A},\eur{B},\eur{C}$ shows
that the first one is equivalent to $\eur{B}$ but worse than $\eur{A}$ and $%
\eur{C}$ . The second algorithm is equivalent to $\eur{A}$ but worse than $%
\eur{B}$ and $\eur{C}$ . Obviously the other algorithms with $P_2$ as the
first predicate (i.e. in the root of tree) are even worse.

\begin{figure}
\begin{center}
\unitlength=1.00mm
\special{em:linewidth 0.4pt}
\linethickness{0.4pt}
\begin{picture}(66.00,24.00)
\put(16.00,21.00){\line(3,-5){4.20}}
\put(12.00,21.00){\line(-3,-4){5.24}}
\put(19.00,9.00){\line(-2,-5){3.99}}
\put(14.00,24.00){\makebox(0,0)[cc]{$X$}}
\put(21.00,11.00){\makebox(0,0)[cc]{$Y$}}
\put(5.00,11.00){\makebox(0,0)[cc]{$\alpha$}}
\put(51.00,11.00){\makebox(0,0)[cc]{, where $X$ is $P_2$ or $P_4$ or $P_7$.}}
\put(22.00,9.00){\line(2,-5){4.00}}
\end{picture}
\caption{\it  The second case}
\label{fig5}
\end{center}
\end{figure}
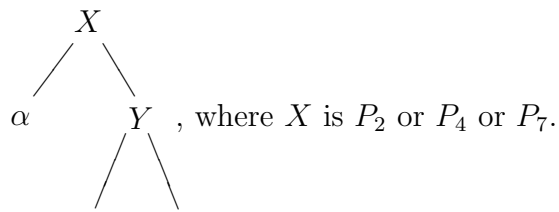

Consider the {\it second case}. Obviously we have to put $P_2$ or $P_4$ or $%
P_7$ in the root if we want to have an image separated (recognized) in one
time unit. So we have all patterns belonging to one image if the first
predicate is true and patterns from the other three images otherwise.
The scheme on [Fig. \ref{fig5}] illustrates that.

Here and below $\alpha ,\beta ,\gamma ,\delta $ mean entire images $\alpha
_i $.

There are three sub cases:
\begin{enumerate}
\item The second predicate ($Y$ on [Fig. \ref{fig5}]) separates 
one entire image;
\item $Y$ separates patterns belonging to half an image;
\item $Y$ doesn't separate either an image or half an image.
\end{enumerate}

\begin{figure}
\begin{center}
\unitlength=1.00mm
\special{em:linewidth 0.4pt}
\linethickness{0.4pt}
\begin{picture}(27.00,31.00)
\put(16.00,28.00){\line(3,-5){4.20}}
\put(12.00,28.00){\line(-3,-4){5.24}}
\put(19.00,16.00){\line(-2,-5){3.99}}
\put(14.00,31.00){\makebox(0,0)[cc]{$X$}}
\put(21.00,18.00){\makebox(0,0)[cc]{$Y$}}
\put(5.00,18.00){\makebox(0,0)[cc]{$\alpha$}}
\put(22.00,16.00){\line(2,-5){4.00}}
\put(13.00,2.00){\makebox(0,0)[cc]{$\frac 12\beta$}}
\put(27.00,2.00){\makebox(0,0)[cc]{$Z$}}
\end{picture}
\end{center}
\caption{\it The second sub case}
\label{subcase22}
\end{figure}
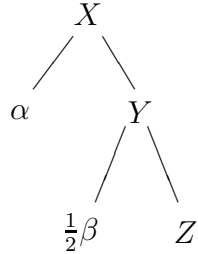

Let us consider the second sub case only (i.e. where $Y$ separates half an
image) [see Fig. \ref{subcase22}].

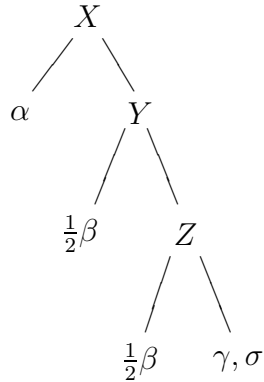
\begin{figure}
\begin{center}
\unitlength=1.00mm
\special{em:linewidth 0.4pt}
\linethickness{0.4pt}
\begin{picture}(34.00,51.00)
\put(16.00,48.00){\line(3,-5){4.20}}
\put(12.00,48.00){\line(-3,-4){5.24}}
\put(19.00,36.00){\line(-2,-5){3.99}}
\put(14.00,51.00){\makebox(0,0)[cc]{$X$}}
\put(21.00,38.00){\makebox(0,0)[cc]{$Y$}}
\put(5.00,38.00){\makebox(0,0)[cc]{$\alpha$}}
\put(22.00,36.00){\line(2,-5){4.00}}
\put(13.00,22.00){\makebox(0,0)[cc]{$\frac 12\beta$}}
\put(27.00,22.00){\makebox(0,0)[cc]{$Z$}}
\put(29.00,19.00){\line(2,-5){4.00}}
\put(25.00,19.00){\line(-1,-3){3.33}}
\put(34.00,5.00){\makebox(0,0)[cc]{$\gamma,\sigma$}}
\put(21.00,5.00){\makebox(0,0)[cc]{$\frac 12 \beta$}}
\end{picture}
\end{center}
\caption{\it  The first case for $Z$}
\label{Z1}
\end{figure}

Now we have to consider possible cases for $Z$. There are two of them. In the
first one $Z$ separates the rest of image $\beta $ [see Fig. \ref{Z1}].

If we try to analyze the situation substituting predicates $P_i$ into $X,Y,Z$
we get [Tab. \ref{dat1}] that shows recognition time for all the cases
except unrealizable ones. A fraction means that half the image is recognized
in time shown in the numerator and the other half is recognized in time
shown in the denominator.

\begin{table}
\begin{center}
\begin{tabular}{|c||c|c|c|}
\hline
image$\backslash$ {\sl case} & 1 & 2 & 3 \\ 
\hline 
$\alpha _0$ & $\frac 23$ & $\geq 4$ & $1$ \\ 
\hline
$\alpha _1$ & $1$ & $\frac 23$ & $\geq 4$ \\ 
\hline
$\alpha _2$ & $\geq 4$ & $1$ & $\frac 23$ \\ 
\hline
$\alpha _3$ & $\geq 4$ & $\geq 4$ & $\geq 4$ \\
\hline
\end{tabular}
\caption{\it  Recognition time for the first case}
\label{dat1}
\end{center}
\end{table}

[Tab. \ref{comp1}] shows the result of comparison those three cases with
algorithms $\eur{A,B,C}$ . Sign '$+$' means that algorithm in the row is
better than algorithm in the column.

\begin{table}
\begin{center}
\begin{tabular}{|c||c|c|c|}
\hline
case & 1 & 2 & 3 \\ 
\hline
$\eur{ A}$ & + & + & + \\ 
\hline
$\eur{ B}$ & + & + & + \\ 
\hline
$\eur{ C}$ & + & + & + \\
\hline
\end{tabular}
\caption{\it  The result of comparison}
\label{comp1}
\end{center}
\end{table}

\begin{figure}
\begin{center}
\unitlength=1.00mm
\special{em:linewidth 0.4pt}
\linethickness{0.4pt}
\begin{picture}(34.00,51.00)
\put(16.00,48.00){\line(3,-5){4.20}}
\put(12.00,48.00){\line(-3,-4){5.24}}
\put(19.00,36.00){\line(-2,-5){3.99}}
\put(14.00,51.00){\makebox(0,0)[cc]{$X$}}
\put(21.00,38.00){\makebox(0,0)[cc]{$Y$}}
\put(5.00,38.00){\makebox(0,0)[cc]{$\alpha$}}
\put(22.00,36.00){\line(2,-5){4.00}}
\put(13.00,22.00){\makebox(0,0)[cc]{$\frac 12\beta$}}
\put(27.00,22.00){\makebox(0,0)[cc]{$Z$}}
\put(29.00,19.00){\line(2,-5){4.00}}
\put(25.00,19.00){\line(-1,-3){3.33}}
\put(34.00,5.00){\makebox(0,0)[cc]{$\frac 12 \beta,\gamma$}}
\put(21.00,5.00){\makebox(0,0)[cc]{$\gamma$}}
\end{picture}
\caption{\it  The second case for $Z$}
\label{Z2}
\end{center}
\end{figure}
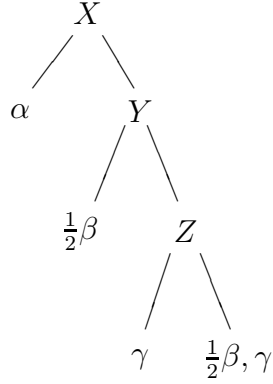

So we see there is no algorithm better than $\eur{A}$ or $\eur{B}$ or $\eur{C%
}$. Let us find out about the second case for $Z$, i.e. when $Z$ separates 
an entire image $\gamma $ [see Fig. \ref{Z2}].

Again we get two tables [Tab. \ref{datandcomp2}] 
which have the same meaning as [Tab. \ref{dat1}, \ref{comp1}].

\begin{table}
\begin{center}
\begin{tabular}{|c||c|c|c|c|c|c|}
\hline
image$\backslash$ {\sl case} & 1 & 2 & 3 & 4 & 5 & 6 \\ 
\hline 
$\alpha _0$ & $1$ & $\frac 24$ & $3$ & $\frac 24$ & $3$ & $1$ \\ 
\hline
$\alpha _1$ & $\frac 24$ & $3$ & $1$ & $1$ & $\frac 24$ & $3$ \\ 
\hline
$\alpha _2$ & $3$ & $1$ & $\frac 24$ & $3$ & $1$ & $\frac 24$ \\ 
\hline
$\alpha _3$ & $\geq 4$ & $\geq 4$ & $\geq 4$ & $\geq 4$ & $\geq 4$ & $\geq 4$ \\ 
\hline
\end{tabular}
\quad\quad
\begin{tabular}{|c||c|c|c|c|c|c|}
\hline
case & 1 & 2 & 3 & 4 & 5 & 6 \\ 
\hline 
$\eur{ A}$ & + & + & + & + & + & + \\
\hline 
$\eur{ B}$ & + & + & + & + & + & + \\ 
\hline
$\eur{ C}$ & + & + & + & + & + & + \\
\hline
\end{tabular}
\caption{\it  Comparison time and result for the second case}
\label{datandcomp2}
\end{center}
\end{table}

It gives us the same result.

We finished the second sub case. The other two sub cases can be investigated
similarly.

The theorem is proved.

\smallskip\ 

There is also a more general theorem on nontransitivity.

{\sl Theorem 2}. {\it Let} $n\geq 3$. {\it Then there are such }$U^f$, $S$, 
{\it and} $P_i$ {\it that there exist} $n$ {\it recognition algorithms} $%
\eur{A}_j$ {\it forming a nontransitive sequence:} $\eur{A}_0\ll \eur{A}%
_1\ll \eur{A}_2\ll \ldots \ll \eur{A}_{n-1}\ll \eur{A}_0$.

{\sl Proof.}

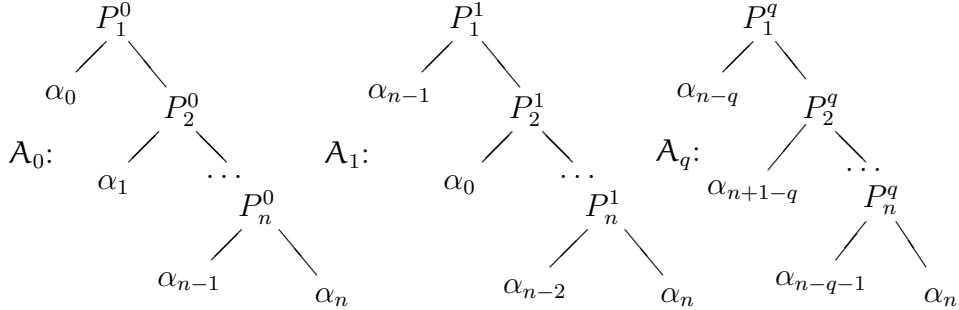
\begin{figure}
\begin{center}
\unitlength=1.00mm
\special{em:linewidth 0.4pt}
\linethickness{0.4pt}
\begin{picture}(121.00,43.00)
\put(13.00,40.00){\line(5,-6){5.00}}
\put(33.00,15.00){\line(5,-6){5.00}}
\put(28.00,15.00){\line(-1,-1){4.00}}
\put(60.00,40.00){\line(5,-6){5.00}}
\put(64.00,28.00){\line(-1,-1){4.00}}
\put(79.00,15.00){\line(5,-6){5.00}}
\put(99.00,40.00){\line(4,-5){4.00}}
\put(115.00,16.00){\line(2,-3){4.00}}
\put(111.00,16.00){\line(-4,-5){4.00}}
\put(103.00,29.00){\line(-5,-6){5.00}}
\put(10.00,40.00){\line(-1,-1){4.00}}
\put(56.00,40.00){\line(-1,-1){4.00}}
\put(96.00,40.00){\line(-1,-1){4.00}}
\put(58.00,43.00){\makebox(0,0)[cc]{$P_1^1$}}
\put(97.00,43.00){\makebox(0,0)[cc]{$P_1^q$}}
\put(11.00,43.00){\makebox(0,0)[cc]{$P_1^0$}}
\put(20.00,31.00){\makebox(0,0)[cc]{$P_2^0$}}
\put(66.00,31.00){\makebox(0,0)[cc]{$P_2^1$}}
\put(105.00,31.00){\makebox(0,0)[cc]{$P_2^q$}}
\put(30.00,18.00){\makebox(0,0)[cc]{$P_n^0$}}
\put(74.00,15.00){\line(-1,-1){5.00}}
\put(76.00,18.00){\makebox(0,0)[cc]{$P_n^1$}}
\put(113.00,19.00){\makebox(0,0)[cc]{$P_n^q$}}
\put(4.00,33.00){\makebox(0,0)[cc]{$\alpha_0$}}
\put(11.00,21.00){\makebox(0,0)[cc]{$\alpha_1$}}
\put(21.00,8.00){\makebox(0,0)[cc]{$\alpha_{n-1}$}}
\put(40.00,6.00){\makebox(0,0)[cc]{$\alpha_n$}}
\put(49.00,33.00){\makebox(0,0)[cc]{$\alpha_{n-1}$}}
\put(57.00,21.00){\makebox(0,0)[cc]{$\alpha_0$}}
\put(67.00,7.00){\makebox(0,0)[cc]{$\alpha_{n-2}$}}
\put(86.00,6.00){\makebox(0,0)[cc]{$\alpha_n$}}
\put(90.00,33.00){\makebox(0,0)[cc]{$\alpha_{n-q}$}}
\put(96.00,20.00){\makebox(0,0)[cc]{$\alpha_{n+1-q}$}}
\put(105.00,8.00){\makebox(0,0)[cc]{$\alpha_{n-q-1}$}}
\put(121.00,6.00){\makebox(0,0)[cc]{$\alpha_n$}}
\put(0.00,25.00){\makebox(0,0)[cc]{$\eur{A}_0$:}}
\put(42.00,25.00){\makebox(0,0)[cc]{$\eur{A}_1$:}}
\put(86.00,25.00){\makebox(0,0)[cc]{$\eur{A}_q$:}}
\put(13.00,24.00){\line(1,1){4.00}}
\put(22.00,28.00){\line(1,-1){4.00}}
\put(68.00,28.00){\line(1,-1){4.00}}
\put(107.00,28.00){\line(1,-1){4.00}}
\put(26.00,22.00){\makebox(0,0)[cc]{$\dots$}}
\put(73.00,22.00){\makebox(0,0)[cc]{$\dots$}}
\put(111.00,23.00){\makebox(0,0)[cc]{$\dots$}}
\end{picture}
\caption{\it  Recognition algorithms $\eur{A}_j$}
\label{aln}
\end{center}
\end{figure}

Define $v_i=\underset{i-1}{\underbrace{0\ldots 0}}1\underset{n-i}{
\underbrace{\frak{B}\ldots \frak{B}}}$, $1\leq i\leq n$. We introduce $n+1$
images as follows: $\alpha _0=\{v_1v_2\ldots v_n\}$, $\alpha
_1=\{v_2v_3\ldots v_nv_1\}$, $\alpha _2=\{v_3v_4\ldots v_nv_1v_2\}$,$\ldots $%
, $\alpha _{n-1}=\{v_nv_1\ldots v_{n-1}\}$, $\alpha _n=\{0\ldots 0\}$. Each
image is a subset of $\frak{B}^{n^2}$. $U^f=\underset{0\leq i\leq n}{%
\bigcup }\alpha _i$. $S=\{\alpha _i\mid 0\leq i\leq n\}$. Then we define the
signs: $P_m^i(x)=a_{ni+m}$, where $x=a_1\ldots a_na_{n+1}\ldots a_{2n}\ldots
\ldots a_{n(n-1)+1}\ldots a_{n^2}$; $0\leq i\leq n-1$, $1\leq m\leq n$.
The above mentioned recognition algorithms can be defined as shown on [Fig.
\ref{aln}].

They have recognition time $T(Y,M)$ shown in [Tab. \ref{timen}].

\begin{table}
\begin{center}
\begin{tabular}{|c||c|c|c|c|}
\hline
$M\backslash Y$ & $\eur{A}_0$ & $\eur{A}_1$ & $\ldots$ & 
$\eur{A}_{n-1}$ \\
\hline 
$\alpha _0$ & 1 & 2 & $\ldots$ & $n$ \\
\hline 
$\alpha _1$ & 2 & 3 & $\ldots$ & $1$ \\ 
\hline
$\vdots$ & $\vdots$ & $\vdots$ & $\ldots$ & $\vdots$ \\ 
\hline
$\alpha _{n-1}$ & $n$ & 1 & $\ldots$ & $n-1$ \\ 
\hline
$\alpha _n$ & $n$ & $n$ & $\ldots$ & $n$ \\
\hline
\end{tabular}
\caption{\it  Recognition time of $\eur{A}_j$}
\label{timen}
\end{center}
\end{table}

\smallskip\ 

The theorem statement follows from that table obviously. Note the example
from theorem 1 can be got from the last theorem when $n=3$.

\end{document}